\documentclass[aps,prd,reprint,groupedaddress,amsmath,amssymb]{revtex4-2}
\usepackage{graphicx}
\usepackage[absolute]{textpos}

\begin{document}

\begin{textblock*}{3in}(2.75in,0.3in)
PHYSICAL REVIEW D \textbf{109}, L061302 (2024)
\end{textblock*}
\begin{textblock*}{2.5in}(5.55in,10.5in)
\copyright~2024 American Physical Society
\end{textblock*}

\title{Spin-down of a pulsar with a yielding crust}
\author{Denis Nikolaevich \surname{Sob'yanin}}
\email{sobyanin@lpi.ru}
\affiliation{I. E. Tamm Division of Theoretical Physics, P. N. Lebedev Physical Institute of the Russian Academy of Sciences, Leninskii Prospekt 53, Moscow 119991, Russia}

\begin{abstract}
In light of the discovery of the long-period radio pulsar PSR J0901-4046, it is interesting to revisit a question about how the magnetized neutron star slows down its rotation. In the case of a weak or liquid outer crust, the mechanism of spin-down becomes unclear because the braking stress cannot then be transmitted from the surface to the main bulk of the star. We show that even if the outer crust does not withstand the surface electromagnetic forces creating the braking torque, the stellar spin-down does not stop, and the matter rearranges so that the necessary electromagnetic forces form in more deep and rigid layers capable of withstanding these forces. The spin-down rate remains the same and corresponds to the transformation of the rotational energy of the neutron star into the energy of the generated relativistic electron-positron plasma. The solid iron surface of ultrastrongly magnetized PSR J0901-4046 appears not to be yielding and withstands the braking stress without breaking.
\end{abstract}

\maketitle

This paper considers one important question that appeared during the studies of the recently discovered interesting radio pulsar PSR J0901-4046 associated with an ultraslowly rotating neutron star \cite{CalebEtal2022}. Traditionally, every neutron star with known values of the rotation period $P$ and its time derivative $\dot{P}$ is placed in the $P$-$\dot{P}$ or $P$-$B$ diagram, where $B$ is the surface magnetic field. Observed radio pulsars in these diagrams are usually located to the left of and above the so-called death line, but this radio pulsar is in a sense an exception. Other ultraslowly rotating objects, GLEAM-X J162759.5-523504.3 and GPM J1839-10, have also been discovered recently \cite{HurleyWalkerEtal2022,HurleyWalkerEtal2023}, but their nature (neutron star vs white dwarf) is still debated \cite{LoebMaoz2022,HakanErkut2022,Katz2022,Tong2023,Konar2023}.

The existence of the death line in these diagrams is explained by the fact that the observed radio emission from a neutron star manifesting itself as a radio pulsar is created by an electron-positron plasma generated in its magnetosphere, but certain conditions are necessary for an efficient plasma generation---a sufficiently fast stellar rotation and a sufficiently strong magnetic field---and this does not always hold. For example, PSR J0901-4046 rotates so slowly ($P\approx75.9$~s) that to explain its activity in the radio band, it is necessary to involve a superstrong magnetic field reaching $3\times10^{16}$~G \cite{Sobyanin2023}. However, in this case, there appears a large discrepancy with the standard estimate of the surface magnetic field, $3.2\times10^{19}[(P/1\text{ s})\dot{P}]^{0.5}$~G \cite{LorimerKramer2005}, used for all neutron stars with a known period and its time derivative but based on the assumption that the neutron star is a magnetized sphere rotating in the vacuum and losing its energy as the magnetic dipole. For the mentioned pulsar $\dot{P}\approx2.25\times10^{-13}$, and the so estimated field appears to be 2 orders smaller and equals $10^{14}$~G. The magnetic field higher than the critical (Schwinger) value $B_\text{cr}=m_e^2c^3/e\hbar\approx4.4\times10^{13}$~G implies the existence of interesting quantum electrodynamic effects \cite{IstominSobyanin2007,Shabad2020,Kim2022,RumyantsevEtal2023}.

The discrepancy is not surprising because the active radio pulsar rotates not in the vacuum but in the plasma generated by it itself, and the situation complicates \cite{BeskinEtal1983,Gruzinov2005,PhilippovSpitkovsky2014}. Even if the neutron star is initially in the vacuum, its magnetosphere is inclined to accumulate charged particles and further transform into the plasma-filled state because of a special configuration of the electromagnetic field \cite{IstominSobyanin2009,IstominSobyanin2010a,IstominSobyanin2010b}. If there is a sufficiently strong magnetic field and an external cosmic background of gamma radiation, high-energy photons entering the magnetosphere from the outside initiate the nonstationary plasma generation and lightning formation, so that the magnetosphere eventually becomes filled with this plasma \cite{IstominSobyanin2011a,IstominSobyanin2011b,IstominSobyanin2011c}. Naturally, the vacuum solution \cite{Deutsch1955} outside the magnetized sphere, implying its magnetic-dipole spin-down \cite{Pacini1968}, is not applicable in this case. This means that the applicability of the vacuum magnetic-dipole model is not obvious even for the neutron star inactive in the radio band.

Nevertheless, the inapplicability of the assumption about the magnetic-dipole spin-down does not mean that the magnetic field at the stellar surface cannot be estimated. The electric potential above the magnetic poles is proportional to the magnetic field, so the energy of the particles being accelerated in the polar gaps of the neutron star and initiating the cascade of production of a relativistic electron-positron plasma contains information about the magnetic field existing there, which is eventually reflected in characteristics of the observed plasma radiation. Detailed consideration of the problem of determination of the surface magnetic field for PSR J0901-4046 is presented in Ref.~\cite{Sobyanin2023}.

While rotating, the neutron star transfers the rotational energy to the particles of the generated plasma and so spins down. The loss of the rotational energy $E$ occurs due to the work of the force of the electric field on the charged particles being accelerated in the polar gap, and this field itself is induced by the stellar rotation. The energy transformation is described by
\begin{equation}
\label{EdotUI}
\dot{E}=VI<0,
\end{equation}
where $V=B R_\text{pc}^2 \cos\alpha/2R_\text{lc}$ is the voltage above the magnetic pole, $B$ is the surface magnetic field, $R_\text{pc}$ is the polar cap radius, $\alpha$ is the magnetic inclination, $R_\text{lc}=c/\Omega$ is the light cylinder radius, $\Omega=2\pi/P$ is the angular frequency, and $I$ is the total uniformly distributed electric current flowing through the polar cap. The signs of the quantities entering Eq.~\eqref{EdotUI} correspond to the north magnetic pole of the neutron star, where the magnetic field is assumed positive, $B>0$. Naturally, all this implies that in order for the neutron star to spin down, some specific physical forces providing the braking torque should act on it.

\begin{figure}
\includegraphics[width=8.6cm]{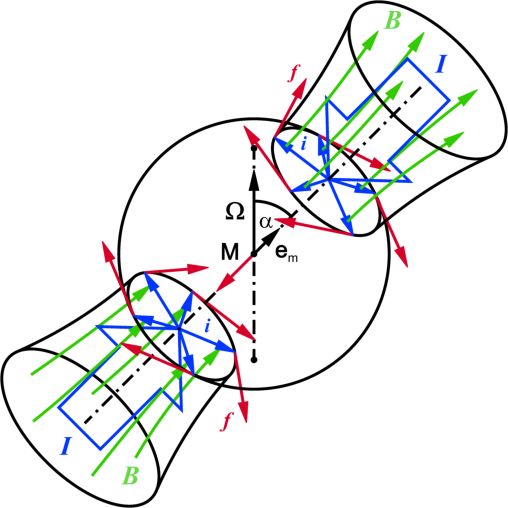}
\caption{Formation of braking torque.}
\label{fig1}
\end{figure}

Let for definiteness $\alpha<90^\circ$ and the electric current $I<0$ be directed to the stellar surface (Fig.~\ref{fig1}). In a magnetic tube of radius $r_\perp<R_\text{pc}$, there is the current $I(r_\perp)=(r_\perp/R_\text{pc})^2 I$, which upon reaching the very surface flows over symmetrically to the sides due to the charge conservation law, and the distribution of currents directed radially from the magnetic axis with the linear density $i(r_\perp)=-I(r_\perp)/2\pi r_\perp=-(r_\perp/2\pi R_\text{pc}^2)I$ forms at the surface. Reaching the edge of the polar cap, the current goes away from the stellar surface and continues to flow in the magnetosphere along the boundary between the regions of open and closed magnetic field lines in the outward direction \cite{ScharlemannWagoner1973,BeskinEtal1993,Kato2017,HuBeloborodov2022}. The moment of Ampere's forces (torque) about the magnetic axis acting on the polar cap is $M_0=\int r_\perp f(r_\perp)dS$, where $f(r_\perp)=i(r_\perp)B/c$ is the stress of Ampere's forces, and the integration is performed over the polar cap surface $r_\perp<R_\text{pc}$, so that $M_0=IBR_\text{pc}^2/4c$. The total torque $M=2M_0$ (there are two polar caps) can be written thus,
\begin{equation}
\label{Mtotal}
M=\frac{I\Psi_\text{pc}}{c},
\end{equation}
where $\Psi_\text{pc}=\Phi_\text{pc}/2\pi$ is the reduced magnetic flux and $\Phi_\text{pc}=B\pi R_\text{pc}^2$ is the total magnetic flux passing through the polar cap; therefore,
\begin{equation}
\label{EdotMOmega}
\dot{E}=\mathbf{M}\cdot\mathbf{\Omega}=M\Omega\cos\alpha,
\end{equation}
where $\mathbf{M}=M\mathbf{e}_\text{m}$ is the vector of torque, $\mathbf{e}_\text{m}$ is the direction of magnetic axis, and $\mathbf{\Omega}$ is the vector of angular velocity. Equation \eqref{EdotMOmega} after substituting Eq.~\eqref{Mtotal} reduces to Eq.~\eqref{EdotUI}.

Now let us turn to one subtlety that we initially intended to consider here. The derivation of Eq.~\eqref{EdotMOmega} is based on the calculation of the moment of Ampere's forces acting directly on the surface of the neutron star. However, mechanical and electrodynamic processes in the stellar envelope are influenced by its phase state and strength. Under certain conditions the outer crust of the neutron star can simply yield to the surface forces and break down. It then becomes completely unclear how the neutron star spins down if its crust cannot transmit the braking stress to the main bulk of the star.

The fact that a magnetized neutron star with such a yielding crust also spins down cannot raise doubts because the derivation of the initial Eq.~\eqref{EdotUI} is based on the energy transformation and does not require consideration of the acting forces and, correspondingly, mechanical characteristics of the crust. However, the forces must exist anyway and act somewhere, seeing that the spin-down is caused precisely by these forces.

Consider a rotating magnetized neutron star with the working polar gaps being the sources of current. Ampere's forces immediately begin acting on the stellar surface, breaking it. The breaking leads to the disappearance of coupling between contiguous volumes of matter and correspondingly, to the appearance of the motion of the outer layers of the crust in the polar caps under the action of Ampere's forces. Such a motion leads to the rearrangement of the distributions of the magnetic field $\mathbf{B}$ and the electric current density $\mathbf{j}$ and consequently, to the rearrangement of the distribution of the acting forces, but it will occur only until the formation of such a configuration of forces that the matter can withstand. In the case of a very weak crust, when any somewhat significant force breaks and moves it, a force-free configuration should form, when all Ampere's forces vanish,
\begin{equation}
\label{forceFree}
\mathbf{j}\times\mathbf{B}=0.
\end{equation}

Assume, for simplicity, that the formed configuration of the fields and currents is axisymmetric but otherwise arbitrary. A given layer of the yielding matter is then passed through by nested magnetic tubes symmetric about the magnetic axis of the neutron star. Every such tube is characterized by some constant value of the magnetic flux $\Phi$ (or reduced magnetic flux $\Psi=\Phi/2\pi$), and the radius of its cross section in general depends on the level at which this tube is intersected by the plane orthogonal to the magnetic axis.

It follows from Eq.~\eqref{forceFree} that the current density is tangent to the lateral surface of the magnetic tube; i.e., charge does not intersect it, and the total current flowing through the tube cross section at one level is equal to the total current flowing through the tube cross section at any other level and is then a characteristic of the tube,
\begin{equation}
\label{Iint}
I(\Psi)=const.
\end{equation}
Note that the property \eqref{Iint} follows from the force-free approximation and does not hold in general \cite{Sobyanin2019}.

Thus, if the matter of the polar caps is yielding, the electric current flowing in the gap and reaching the surface does not flow over it but goes inside the neutron star along the magnetic tubes. This will occur while the force-free approximation \eqref{forceFree} holds, i.e., until the current reaches deep matter layers capable of withstanding Ampere's forces and not breaking. After this, the force-free approximation becomes inapplicable, and the electric current begins to flow over the solid matter radially from the magnetic axis analogously to the already considered situation of the currents flowing over the solid stellar surface. Reaching the outer magnetic tube, characterized by the reduced magnetic flux~$\Psi_\text{pc}$, the current goes along its lateral surface back to the stellar surface and then flows in the magnetosphere along the surface embracing the bundle of open magnetic field lines.

Let us calculate what torque Ampere's forces will give in this case. It is better to characterize the distance from the magnetic axis to a given point not by the radial coordinate but by the value of the reduced magnetic flux $\Psi$ passing through the magnetic tube on the lateral surface of which the point lies. Integrating over the boundary between the solid and yielding matter inside the outer magnetic tube, we obtain the total moment of acting forces for two polar caps,
\begin{equation}
\label{Mtotal2}
M=\frac{2}{c}\int_0^{\Psi_\text{pc}}I(\Psi)d\Psi.
\end{equation}
Since $\Psi$ and $I(\Psi)$ are conserved during motion along magnetic tubes, so is the torque,
\begin{equation}
\label{MtotalInt}
M=const,
\end{equation}
and the integral may be taken over any cross section of the outer magnetic tube by the plane orthogonal to the magnetic axis and lying in the yielding matter, in particular over the polar cap at the stellar surface. Since we assumed that the electric current and longitudinal magnetic field are distributed uniformly when initially considering the currents flowing over the stellar surface, then $I(\Psi)=I\Psi/\Psi_\text{pc}$. Substituting this relation into Eq.~\eqref{Mtotal2}, we again arrive at Eq.~\eqref{Mtotal}. This means that the spin-down of a magnetized neutron star is again expressed by Eq.~\eqref{EdotUI}.

During the derivation of Eqs.~\eqref{EdotMOmega} and \eqref{MtotalInt}, the uniformity of the distribution of the longitudinal electric current over the polar cap is not used, so Eq.~\eqref{EdotUI} for the energy loss can be generalized to the case of a nonuniform distribution of the current. From dynamic Eqs.~\eqref{EdotMOmega} and \eqref{Mtotal2}, we obtain a more complex expression,
\begin{equation}
\label{EdotUIcomplex}
\dot{E}=\frac{2\cos\alpha}{R_\text{lc}}\int_0^{\Psi_\text{pc}}I(\Psi)d\Psi.
\end{equation}

It is interesting to obtain Eq.~\eqref{EdotUIcomplex} not from dynamic but from energy considerations. Over the polar cap through a ring with the radius determined by $\Psi$ and the thickness determined by $d\Psi$, there flows the electric current $dI(\Psi)=I'(\Psi)d\Psi$. The accelerating voltage is distributed over the polar cap parabolically and for this ring is equal to
\begin{equation}
\label{VPsi}
V(\Psi)=V\biggl(1-\frac{\Psi}{\Psi_\text{pc}}\biggr),
\end{equation}
where the voltage at the magnetic pole is conveniently written as $V=\Psi_\text{pc}\cos\alpha/R_\text{lc}$. The electromagnetic power released by the neutron star in its two polar caps is then
\begin{equation}
\label{EdotUIcomplex2}
\dot{E}=2\int_0^{\Psi_\text{pc}}V(\Psi)dI(\Psi).
\end{equation}
Integrating in Eq.~\eqref{EdotUIcomplex2} by parts, we arrive at Eq.~\eqref{EdotUIcomplex}.

The equality of the work of Ampere's forces on the neutron star, which brake its rotation, and the work of the longitudinal electric field on the current, which accelerates charged particles, indicates the transformation of the rotational energy of the neutron star into the energy of the generated electron-positron plasma. This can also be verified otherwise---considering the balance of Poynting fluxes on the magnetic poles. Let us calculate the flux $F_\text{em}^\text{surf}$ of electromagnetic energy going directly from the surface of the neutron star and the flux $F_\text{em}^\text{out}$ of electromagnetic energy above the gaps. The electric field in the magnetosphere has the form $\mathbf{E}=\mathbf{E}_0-\nabla\varphi$ \cite{Mestel1971}, where $\mathbf{E}_0=-(1/c)\mathbf{v}\times\mathbf{B}$, $\mathbf{v}=\mathbf{\Omega}\times\mathbf{r}$, $\mathbf{r}$ is the radius vector, and $\varphi\rightarrow V(\Psi)$ above the gaps. Since $\nabla\varphi$ is normal to the stellar surface, we have
\begin{equation}
\label{F0}
F_\text{em}^\text{surf}=\frac{c}{4\pi}\int(\mathbf{E}_0\times\mathbf{B})\cdot d\mathbf{S}.
\end{equation}
Equation \eqref{F0}, with taking account of the relations $B_\phi=2I(\Psi)/cr_\perp$ for the azimuthal magnetic component and $d\Phi=\mathbf{B}\cdot d\mathbf{S}$ for the magnetic flux, transforms into
\begin{equation}
\label{F01}
F_\text{em}^\text{surf}=-\mathbf{M}\cdot\mathbf{\Omega}=-\dot{E},
\end{equation}
where $M$ is determined by Eq.~\eqref{Mtotal2}. Equation \eqref{F01} reflects that the flux of electromagnetic energy leaving the surface of the neutron star is equal to the rate of rotational energy loss.

The flux of electromagnetic energy above the gaps can be represented in the form
$F_\text{em}^\text{out}=F_\text{em}^\text{surf}-F_\text{em}^\text{gap}$, where
\begin{equation}
\label{Fgap}
F_\text{em}^\text{gap}=\frac{c}{4\pi}\int(\nabla V(\Psi)\times\mathbf{B})\cdot d\mathbf{S},
\end{equation}
with the integration in Eqs.~\eqref{F0} and \eqref{Fgap} now not over the stellar surface but over the spherical surface above the polar gaps. The integration in Eq.~\eqref{F0} over the new surface leads to the same result \eqref{F01} for the conserving $I(\Psi)$, so $F_\text{em}^\text{gap}$ corresponds to the flux of electromagnetic energy being absorbed in the polar gaps. Equation \eqref{Fgap} becomes $F_\text{em}^\text{gap}=(c/4\pi)\int(B_\phi/r)(\partial V(\Psi)/\partial\theta)dS$, where $\theta$ is the polar angle with respect to~$\mathbf{e}_\text{m}$. From $B_\phi=2I(\Psi)/cr_\perp$ and $dS=2\pi r r_\perp d\theta$, we have for two polar caps
\begin{equation}
\label{Fgap2}
F_\text{em}^\text{gap}=2\int_0^{\Psi_\text{pc}} I(\Psi) dV(\Psi).
\end{equation}
Substituting Eq.~\eqref{VPsi} into Eq.~\eqref{Fgap2} and comparing to Eq.~\eqref{EdotUIcomplex} yields $F_\text{em}^\text{gap}=-\dot{E}$ and $F_\text{em}^\text{out}=0$, which reflects the absorption of the Poynting flux in the polar gaps and its transformation into the energy of accelerated electrons and positrons. Naturally, the plasma energy can immediately or further transform into electromagnetic radiation again, such as curvature gamma rays, thermal x rays from the surface heated by accelerated particles, or plasma radio emission.

Finally, let us find out which of the two cases considered relates to PSR J0901-4046 and whether its crust is yielding. From the results of Ref.~\cite{ParmarEtal2023}, we conclude that in a strong magnetic field of $B\approx3.2\times10^{16}\text{ G}=3.2$~TT, the pulsar surface is solid, has a high baryon density of $\rho\sim10^{-6}\text{ fm}^{-3}$ (Table~1, we take $B=1000 B_\text{cr}$ to estimate the density), and still remains iron [Fig.~1(a)]. The shear modulus of a crystalline crust can be estimated as $\mu\sim0.1n(Ze)^2/a$ \cite{OgataIchimaru1990}, where $Ze$ is the charge of one ion of the crystal, $a=(3/4\pi n)^{1/3}$ is the radius of the sphere with the volume occupied by one ion, $n=\rho/A$ is the ion density, and $A$ is the mass number. For iron $Z=26$ and $A=56$, so we obtain $\mu\sim10^{25}\text{ Ba}=1$~YPa. The electric current flowing in the polar cap is $I\approx56$~MA \cite{Sobyanin2023}, so at the edge of the polar cap, there is a linear current density of $i=I/2\pi R_\text{pc}\approx0.53\text{ MA}/\text{m}$, where the polar cap radius is $R_\text{pc}\approx17$~m. Ampere's forces create on the solid surface a shear stress of $\tau=iB\approx1.7$~EPa (in SI), causing a shear strain of $\gamma=\tau/\mu\sim10^{-6}$. The critical strain under which the crust breaks can be estimated as $\gamma_\text{cr}\sim0.1$ \cite{HorowitzKadau2009,HoffmanHeyl2012}. Since $\gamma\ll\gamma_\text{cr}$, the crust of PSR J0901-4046 withstands the braking stress and does not break, thus being not yielding.

In conclusion, even if the outer crust of the neutron star does not withstand Ampere's forces trying to brake the star but instead simply breaking the crust and causing the motion of its matter initiating reconfiguration of the magnetic field, in such a matter, there eventually forms a force-free configuration of the magnetic field letting in the electric current generated above the polar caps inward the neutron star until the current reaches the layers of matter capable of withstanding these Ampere's forces. The resulting braking torque appears to be equal to the braking torque acting on the neutron star with a solid and rigid surface. The results of this paper are applicable to any neutron star with active polar caps.

\end{document}